\documentclass{article}

\usepackage{amsmath}
\usepackage{amssymb}

\begin{document}

\title{The Essence of Quantum Theory for Computers}
\author{\small W.C. Parke, Prof. Emeritus of Physics, GWU}

\maketitle

\begin{abstract}
Quantum computers take advantage of interfering quantum alternatives in order to handle problems that might be too time consuming with algorithms based on classical logic.
Developing quantum computers requires new ways of thinking beyond those in the familiar classical world.
To help in this thinking, we give a description of the foundational ideas that hold in all of our successful physical models, including quantum theory.  Our 
emphasis will be on the proper interpretation of our theories, and not just their statements.  Our tact will be to build on the concept of information, which lies central to the operation of not just computers, but the Universe. For application to quantum computing, the essence of quantum theory is given, together with special precautions and limitations.
\end{abstract}

\section{Introduction}
Having a grasp on the ideas behind a theory helps to apply it correctly, to understand its limitations, and to  generate new ideas. Getting a firm hold on quantum theory is not an easy task, because our experiences and even our genetic predispositions have been developed in a world in which quantum effects are largely washed out.\footnote{Although these days, macroscopic quantum effects can be seen in the actions  of lasers and of quantum fluids.} Remarkably, our predilection for finding logic behind the behavior of what we observe,\footnote{Our curiosity is enhanced by genetic selection, as there is advantage to being able to
make sense of what goes on around us, so that
we can anticipate what might happen next.} including that of electrons and atoms, has led us to quantum theory, a description of nature that is hard for us to conceptualize, but is logical, accurate, and explains a wide variety of phenomena with only a few statements and input.

As background to quantum theory and quantum computing,
an attempt is made here to give the primitive notions and essential observations that underlie current physical theories, so
that foundational ideas are explicit, and a common language is established. In our description, information
storage and transfer is made central.\footnote{Traditionally, energy transfer is used to characterize interactions in current theories.  However, the concept of energy is several steps removed from more basic ideas.  Moreover, information processing is not only the purpose of computers, but also lies underneath all natural processes.}  A short description
of quantum theory follows, and then applied to quantum computing, focusing on what the theory says, and particularly does not say, in areas where conceptual difficulties have arisen.

\section{Physical theory and reality}
A physical theory is a logical model capable of making predictions of what we observe.  It is judged by its accuracy in matching measurements, and by its economy, i.e. whether the proposed theory has only a few relationships and input data needed for its ability to explain observations over a wide realm.\footnote{In information theory terms, the information contained in the independent
data explained by a theory should be much larger than the information needed to
express the theory.}

We should not, however, become too enamored with the auxiliary structures
within a successful theory.
Just as it is possible to transform, isomorphically, a logical structure
into an equivalent one involving distinctly different relationships and
symbols, it is also possible to so transform a physical theory. A
good example is the transformation of Maxwellian electrodynamics into
an action-at-a-distance form. The transformed theory, invented by Wheeler and Feynman,\footnote{John Archibald Wheeler and Richard Phillips Feynman,  
``Classical Electrodynamics in Terms of Direct Interparticle Action'', {\it Reviews of Modern Physics},
{\bf 21},\,pp.\,425-433 (1949).}
no longer contains electric or magnetic fields. Even so, it
makes the same predictions as Maxwell's theory.\footnote{We generally use Maxwell's theory 
to solve electrodynamics problems because the Wheeler-Feynman 
theory is a more complicated mathematical
system.}  A lesson from this example and others is that one should not endow  physical
meaning to all the symbols and relationships in a theory.  Electric fields do not `exist' in nature.
They exist as symbols on paper and in our minds.  But Maxwell's theory does make definite statements 
about observations using the electric field concept.
Only those points in the
theory that are stated as predictions can be connected to nature. In
quantum theory, wave functions are clearly not physical; in general, 
they are complex numbers. They can also be transformed away
in alternate but equivalent theories.\footnote{For example, Werner Heisenberg's formulation of quantum theory, shown by P.A.M. Dirac to be completely equivalent to Erwin Schr\"{o}dinger's, uses no wave functions. Neither do various so-called hydrodynamic formulations, such as that of Erwin Madelung in ``Quantentheorie in Hydrodynamischer Form'',
{\it Z. Phys.}\,{\bf 40},\,pp.\,322-326 (1927).} Rather, one should think of the symbols and relationships in a theory as {\it tools for making predictions}. Predictions are the
touchstones in the theory. All else is ancillary. 

Here is another caution: 
Predictions of pure counts are testable as either true or false, but predictions of
continuous values will never be
proved to match nature exactly, since our measuring instruments are finite.  Theories which take space as continuous implicitly
do so only down to the scale permitted by
our instruments.  There should be no implication that even continuity exists at finer scales.

Our best physical theory so far is the so-called `Standard Model',\footnote{For a personal perspective in the development of the Standard Model, see Steven Weinberg's article,
``The Making of the Standard Model'', {\it Eur.Phys.J.}\,{\bf C34},\,pp.\,5-13 (2004).}
which describes, with quantum field theory, all of the interactions yet detected,
except for gravity.
The Standard Model has made remarkable and now verified predictions and agrees with the most
precise of measurements made to one part in a trillion. Even
so, the theory is not tight, having many unexplained interaction strengths and masses.  
We expect new theories will give a deeper and simpler explanation of particles, of their interactions, and of the yet unexplored regions in nature.

In the next section, a set of tentative propositions and observations underlying all physical theories is proposed, building toward the foundations of quantum theory and application to quantum computing.  Information storage and transfer will be seen to be fundamental to natural processes.

\section{Basic properties of physical systems}

The natural world is divisible into a collection of observable subsystems.  Each observable
subsystem will be referred to as a {\bf physical system}.  If a physical
system can be further divided, the parts may be called `components' of the system.  The number of divisions may reach a limit.  

A physical system can store {\bf information}, taken to be an additive quantity which grows
with the number of distinct ways
that the system may be configured
under given physical constraints. The number of ways is called
the system's `{\bf multiplicity}', $W$.\footnote{One of the many remarkable
implications of quantum theory is that the count $W$ can be performed
over a denumerable number of quantum states of a system.}
To be additive across independent systems, the information $I$ in
a system must be proportional to $\ln{W}$.\footnote{If there were two independent systems of multiplicity $W_1$ and $W_2$, then the multiplicity of 
both together would be $W_1W_2$.  The condition $f(W_1W_2)=f(W_1)+f(W_2)$ makes $f(W)$ proportional to $ln(W)$.}  With $I=\ln_2{W}$, the information is given in `bits'.\footnote{If a given system subject to physical constraints cannot be re-configured, then that system has only
one bit of information. If the system has two possible configurations, its reading transmits one bit of information, the equivalent of a yes or a no, but no more, and so forth.}
 If the multiplicity $W$ of a system decreases, we say the system has become more `{\bf ordered}'.\footnote{In the late nineteenth century, Ludwig Boltzmann introduced the number $W$ (`Wahrscheinlichkeit'), connecting it to the disorder (Clausius' `entropy', $S$) of
a system with $S=k \ln{W}$, where $k$ is Boltzmann's constant.  Le{\'o} Szil{\'a}rd showed that each bit of information we gather from
a system and discard necessarily requires an increase in entropy
of at least $k\ln{2}$. (``On the 
Decrease of Entropy in a Thermodynamic System by the Intervention
of Intelligent Beings'', {\it Zeitschrift f{\"u}r Physik,}\,{\bf 53},\,pp.\,840-856 (1929).) 
Claude Shannon 
developed the formalism of information theory, including information
transfer in the presence of noise. (Shannon,\,C.E., ``A Mathematical Theory of Communication'', {\it Bell System Technical Journal},\,{\bf 27},\,pp..\,379-423 \& 623-656, July \& October, 1948).}  

An {\bf interaction} between two physical systems, by definition, exchanges {\it information} between them.  
An {\bf open physical system} can interact with other systems.  
{\bf Observation} is made by allowing two physical systems to interact, one
of which is prepared as a measuring instrument.  A {\bf measuring instrument}
is a physical system whose information
gathered from an observed system is capable of being copied with a relatively high assurance. 
The copy will act as a {\bf record} of the observation.  Statements about a physical system 
are verified only by observations.\footnote{
This grounding is particularly poignant
in quantum theory, wherein a quantum system is described by a set of interfering possible states
for each observable, with only one such state realized by observation.} 
A statement about a physical system is {\bf predictive} if it relates a number of observations of that system.  
A physical system is {\bf isolatable} if the measurable effect of all interactions
with other external systems can be made arbitrarily {\it small}.\footnote{We will
use the term `small' for a quantity which has the property that if made
smaller, there would be no significant effect.}
An isolated physical system is said to be `closed' when external interactions which might influence
the results of intended measurements of that system are negligible.

If a set of observations of a system is found to repeat,
that system can act as a {\bf clock}, with {\bf time} {\it defined} and measured
by the number of repeats, each smallest repeating cycle called a {\bf period} of the clock.
If a large set of independent periodic systems, prepared in the same way, 
are found to consistently have the same number $N$ of periods, these clocks 
are said to be `good' to a precision of at least one part in $1/N$.

The {\bf distance} between two interacting physical systems is defined, up
to a selected constant factor, to be the minimum time needed for an observable change in one of those two systems to cause an observable effect in the other. 
{\bf Space} is
defined to be the set of available distances between all systems.
Two systems with a finite distance
between them are said to be spatially separated. 
If one isolatable
system can be spatially separated from all others, it is {\bf localizable}.\footnote{Defining the localizability of zero-mass particles with spin greater than $1/2$ (in units of Planck's constant over $2\pi$), such as the photon, is tricky. For a definition, and references back to Wolfgang Pauli, see Margaret Hawton, ``Photon position operator with commuting component'', {\it Phys.Rev.} {\bf A 59 (2)}, pp.954-959 (1999).}
If $N$ localizable systems can be
spatially separated from each other by the same distance, then space has at least $N-1$ spatial dimensions. A system localizable in each spatial 
dimension can be referred to as a {\bf body}.
The {\bf spatial coordinates} of a body are the minimal set of 
numbers that uniquely determine a definable {\bf location} within the body.  These coordinates
are measured by one observer relative to
an `origin', a location used by that observer to coordinate a set of bodies.
In an $N$ dimensional
space, a complete set of such coordinates for one location is denoted $\{x^1, x^2,\cdots, x^N\}$. 
An {\bf event}, $\{x^0, x^1, x^2,\cdots, x^N\}$,  specifies when and where an observation has occurred. 

A {\bf frame of reference} characterizes how one observer records
events.
If the spatial separation between two bodies changes with the observer's  time, we say they
have relative motion. Bodies with no average motion relative to the
observer are said to be stationary.
The {\bf velocity} of a body is its spatial change per unit observer's time along each of the independent 
spatial directions, and the {\bf acceleration} is the change of velocity per unit time,
each measurement made in a single frame of reference.

A {\bf particle}  is a localizable physical system with some identifiable intrinsic characteristics,  i.e. quantities that are independent of how the observer measures them.
A {\bf fundamental particle}
is a particle that suffers no measurable change its intrinsic information
even after engaging in all available interactions or after long times.  A `free particle' is a particle 
whose interactions with other systems can be neglected.

Recording a {\bf complete set of observables} in a system determines (to
the degree possible) the 
information present in that system at the time of measurement and before any further interaction
with the system.  The selection of
observables is made such that the measurement of any one does not change
the result that would be found for the measurement of any other in the selected set.
Those observables that are time independent are called {\bf conserved}.

The `dynamics' of a physical system, i.e. a description of how interacting subsystems change over time, follows logical predictive schemes which reveal {\it cause and effect}.  These schemes are
most easily tested using isolatable  `simple systems', i.e. those
with only a few discernible component subsystems and low information content.
So far, all physical systems can be described by the interactions of fundamental particles in space-time.

Systems with many interacting components, called
`complex systems' or `macrosystems', have been successfully described 
when those components can be tracked, or when statistical likelihood
arguments become meaningful.  
Systematics in the behavior of complex systems make global properties referred to as {\bf emergent relationships}. Those of thermodynamics and statistical mechanics are examples.  Rules for optimal dynamics in biosystems\footnote{A `biosystem' is a physical system
whose activities support life.  A life system is one which is capable of self-replication by
interactions with external systems, using information stored within the life system.} form others.  

Some systems, through the mutual interactions of their particles, will form bound bodies, i.e.
systems that retain their localized character provided external
interactions are sufficiently weak.  
A {\bf confined} system is one which, when initially localized in a certain volume with zero average velocity, and then
left alone, will have a non-zero lower bound on the probability of being
found in the initial volume later in time.
The ability to create bound systems gives
preference to the evolution of differentiated systems and 
to condensation into locally ordered subsystems.
With a sufficient variety of
particles and interactions, the evolution of complexity in open subsystems is natural,\footnote{Lars Onsager, ``Reciprocal relations in irreversible processes'', I \& II, 
{\it Phys.Rev.}\,{\bf 37}\,\&\,{\bf 38}, pp.\,405-426, 2265-2279 (1931); Ilya Prigogine, {\it Introduction to Thermodynamics of Irreversible Processes},  New York: Interscience (1955).} including
the evolution of life.

The {\bf Universe} is defined as the collection of everything that can be observed.

\section{Space-time as background to quantum theory}

The Universe appears to have existed in a finite number of current clock
periods, and the volume of our Universe apparently is also finite.  
There is a limit to the greatest separation between bodies.
The dimension of our space is at least three.\footnote{Three dimensions is also the
minimum dimension needed to build a computer or brain having more than four devices with mutual connections. At present, there is no evidence for higher dimensions than three.  The strong experimental support of our conservation laws in three dimensions suggests that if higher dimensions of space existed, matter and energy would have had extreme difficulty passing into or out of it.}
The
distance between widely separated bodies has been growing relative to the size 
of the smallest bodies since time started.

Observation shows that, to a good approximation, there exist {\bf inertial frames} in which 
an isolated body nearby and initially stationary relative to the observer will continue to be nearly stationary. We will use the term `inertial observer' for an observer
in an inertial frame.
In inertial frames, an interaction experienced by one body can
always be associated
with the effect of other local bodies. 
At small scales, the relationships between local events
can be expressed in a form that is
independent of the observer's position, orientation, or motion relative to
the events. This is the grand `{\bf Principle of Relativity}'.\footnote{Radiation
from distant galaxies and radiation left over from the hot big bang do 
establish a unique frame of reference, but these are taken as part of the initial conditions
in dynamics and so do vitiate the relativity principle. In our Universe, the
residual effects of these initial conditions on present observations of local events are often small.}
The Principle of Relativity allows for the existence of a finite {\bf universal limiting speed} for all bodies.\footnote{As demonstrated by
Henri Poincar\'e in ``L'\'etat actuel et l'avenir de la physique mathématique'', St.\ Louis Conference,
{\it Bulletin des sciences math\'ematiques}\ {\bf 28}, pp.302-324 (1904). 
Einstein's second postulate, the constancy of the
speed of light, is not needed.  Relativity alone, under reasonable assumptions about how events are measured
in close by inertial frames with relative motion, initially aligned, allows
only one relationship between their space-time coordinates.  That relationship is the Lorentz transformation, containing a fixed universal speed called $c$.  Explicitly, if the second inertial frame moves
at a speed $v$ away along the positive $x$-axis of the first, then
$x_2=(x_1-vt_1)/\sqrt{1-(v/c)^2}$, $y_2=y_1$, $z_2=z_1$, $t_2=(t_1-(v/c^2)x_1)/\sqrt{1-(v/c)^2}$. 
The Galilean transformation is approached when the universal speed in the Lorentz transformation is taken much larger than the relative speeds
of the observed bodies. This makes $t_2\cong t_1$, so that time becomes
universal in this limit.}
Examination shows that our Universe
has a finite limiting speed.
To the precision of current measurement,  the interactions due to electromagnetism and gravity carry information between bodies at the universal speed $c$.

In Relativity, one observer's measure of spatial separation between
two bodies is related to a combination of space and time coordinates
of another observer moving relative to the first.  This makes the
concept of space and time inseparable, and gives utility
to the idea of a {\it four-vector} using
the coordinates in space and time for a pair of close by events, in the form $\{dx^{\mu}\}=\{dx^0,dx^1,dx^2,dx^3\}$, where $x^0\equiv ct$.  Any
other ordered set of four quantities forms a four-vector if they
transform by coordinate transformations just like $\{dx^{\mu}\}$ do. 

Relativity makes the small interval between two events,  $ds=\sqrt{g_{\mu \nu} dx^{\mu}dx^{\nu}}$ invariant,\footnote{By convention, repeated indices, one upper, one lower, should be summed from 0 to 3.} i.e. independent of the observer's frame of reference. The set of quantities $\{g_{\mu \nu}\}$ form what is called the {\bf metric tensor}.  Each infinitesimal space-time region within any inertial frame can be covered by an orthogonal coordinate grid, 
so that the metric tensor
is well approximated by $\{g_{\mu \nu}\}\approx diag\{1,-1,-1,-1\}$.  A vector `dual' to
$\{dx^{\mu}\}$ can be defined by $dx_{\mu}\equiv g_{\mu\nu}dx^{\nu}$, so that 
$dx_{\mu}dx^{\mu}$ is a `scalar', i.e. a number who value is
independent of the frame of reference of the observer.  
The sum $A_{\mu}B^{\mu}$
defines the scalar product of the two vectors, and the length of $A$ is
$\sqrt{A_{\mu}A^{\mu}}$.  
 An important example of a four-vector is a particle's four-momentum, $\{p^{\mu}\}$, with $c p^0$ being the energy of the particle and $\vec{p}$ its spatial momentum.  The length of $\{p^{\mu}\}/c$ is the mass of the particle.\footnote{The energy and momentum of a system are best defined, in our successful theories, through the generators of time and space translations, with a scale determined by gravity.  These ideas will be 
presented shortly in the context of Noether's Theorem and Einstein's General Relativity Theory.}

A general coordinate transformation between frames of reference, $x'=f(x)$, becomes a `Poincar\'e transformation'
when $x'^{\mu}=a^{\mu}_{\ \nu}x^{\nu}+b^{\mu}$  and the coefficients $\{a^{\mu}_{\ \nu}\}$ satisfy $g_{\mu\nu}a^{\mu}_{\ \kappa}
a^{\nu}_{\ \lambda}=g_{\kappa\lambda}$.\footnote{Note that
this relation makes the metric components an `invariant tensor',
in that the components take the same values after a coordinate transformation.}  Rotations, Lorentz transformations, and displacements are included.  The set $\{a^{\mu}_{\ \nu}, b^{\kappa}\}$ forms  the so-called `Poincar\'e group', with the product rule $\{{a''}^{\mu}_{\ \nu}, b''^{\kappa}\}=\{{a'}^{\mu}_{\ \lambda}a^{\lambda}_{\ \nu}, ({a'}^{\mu}_{\ \lambda}b^{\lambda}+ b'^{\kappa})\}$.\footnote{Reflections are excluded by imposing $det\left|a\right|=1$.  Then the transformations are called `proper'.}

A body initially stationary in an inertial frame, but acted on by one other body some distance away, will accelerate. If a
duplicate of the first body is weakly bound to the first, and the
experiment repeated, then the acceleration of the pair will be 
half the rate of the single one. We say the pair has twice the `{\bf inertial mass}'
of the single body.  
The inertial mass of a particle 
is an intrinsic property.   

The observation of the effects on the motion of bodies due to the
acceleration of the observer's 
frame with respect to an inertial frame
is locally indistinguishable from the effects of
gravity. This is Einstein's {\bf Equivalence Principle}.
Einstein's Equivalence Principle
makes inertial mass the same as {\bf `gravitational mass'}, which is the 
intrinsic property of a body that determines the strength of its gravitational
influence on nearby systems.\footnote{
The Equivalence Principle also means that mass $m$ can be measured in
distance units by giving $Gm/c^2$, where $G$ is Newton's gravitational constant that determines the strength of gravity.}
The mass of any localized system (including the equivalent mass of any associated localized field energy) can be
measured by using the gravitational pull that system creates on a distant mass.

The Equivalence Principle, together with the Principle of Relativity,  
requires that the distance
measure of space-time in the presence of a gravitating body be non-Euclidean, i.e. there will
be intrinsic curvature to the space-time around a body with mass, and 
the metric tensor $\{g_{\mu \nu}\}$ can no longer be 
transformed by a coordinate choice to the form $\{g_{\mu \nu}\}=diag\{1,-1,-1,-1\}$ in any
finite region of the space near the body.  
However, even in the presence of mass, 
inertial observers will still find an approximate flat metric 
in their infinitesimal neighborhood.  

Einstein showed that the effects of gravity due to masses could be
found from conditions on the Riemannian curvature of space-time.
Curvature can be characterized by
the behavior of vectors as they are moved from one
point to another across space.
Infinitesimal changes in any vector that are observed while transporting
that vector along a path define the
`covariant derivative': 
$D_{\kappa}A^{\mu}=\partial_{\kappa} A^{\mu}-\Gamma^{\mu}_{\nu\kappa} 
A^{\nu}.$  The
changes due to the underlying geometry come from the `connections' $\Gamma^{\mu}_{\nu\kappa}$
in the space. In Riemannian geometry, the
connections are determined by gradients of the metric tensor.\footnote{In the
form 
$g_{\mu\lambda}\Gamma^{\lambda}_{\nu\kappa}=(1/2)(\partial_{\kappa} g_{\mu\nu}+\partial_{\nu}g_{\mu\lambda}-\partial_{\mu} g_{\nu\kappa}).$} 
The vector $A^{\mu}(x_0)-\int_{x_0}^{x}\Gamma^{\mu}_{\nu\kappa} 
A^{\nu}dx^{\kappa}$ is said to be the components of the `parallel transport' 
of the original vector at $x_0$ along a particular path to $x$.
The change $\delta_{\square} A^{\mu}$ in the components of any vector field, $A^{\mu}(x)$, by carrying the vector in
parallel transport
around an infinitesimal closed loop, must be proportional to the area of the loop
and the size of the original vector field.  The proportionality 
constants in each small patch of space-time
defines the curvature tensor
$\{{\mathcal R}^{\mu}_{\ \nu \kappa \lambda}\}$ in that patch, to wit:
$\delta_{\square} A^{\mu}={\mathcal R}^{\mu}_{\ \nu \kappa \lambda}A^{\nu}dx^{\kappa}dy^{\lambda}$, where
the loop is given orthogonal sides $dx^{\mu}$ and $dy^{\mu}$.
 
Einstein's General Theory of Relativity 
\footnote{A. Einstein, ``Die Grundlage der allgemeinen Relativit{\"a}tstheorie'', {\it Annalen der Physik}\,{\bf 49},\,pp.\,50-205 (1916).}
is the simplest of a class of theories that incorporate the
Equivalence Principle and the Principle of Relativity.\footnote{More general theories can be constructed using higher derivatives of the metric tensor in the field equations than the second.}   Einstein discovered
that in empty space, the condition on the metric curvature tensor\footnote{The metric curvature tensor $\{R^{\mu}_{\ \nu \mu \lambda}\}$ is that part of the local curvature tensor $\{{\mathcal R}^{\mu}_{\ \nu \kappa \lambda}\}$ due solely
to changes in the metric across space-time.} given by 
$R^{\mu}_{\ \kappa \mu \lambda}=0$ 
numerically predicts: Newtonian gravitational fields when
the effects of gravity differ little from flat space; The size of the extra perihelion precession
of Mercury's orbit;  The amount of the gravitational deflection of light, and; The interval for the slowing of clocks in a gravitational field.  All
these and more have been confirmed to the precision of current instruments.\footnote{Calculations of position on Earth using Global Positioning Satellites at height $h$ and speed $v$ over an Earth of mass $M_E$ and radius $R_E$, have Special Relativity corrections included to order $v^2/c^2$ for the relativistic Dopper shift and General Relativity corrections included to order $GM_E h /(c^2R_E^2)$ for clock slowing in a gravitational field.  Without these, errors in positions would be unacceptable!}

In both the Special and the General Theory of Relativity, time is not universal. If two good clocks are synchronized in one frame of
reference, and one is set in motion relative
to the other, they may differ in the number of periods each had when they are brought back together.\footnote{
This leads to the `Twin Paradox', that one twin can end up
younger than the other, yet each sees the other move away and
then come back.  The resolution came from
Einstein using his General Theory of Relativity.
The difference in the time elapsed by the clocks will be the difference between the values of
$\int\sqrt{\left|g_{\mu\nu}dx^{\mu}dx^{\nu}\right|}/c$, integrated along the path of each clock from the common starting point to the common endpoint.}

In General Relativity, bodies acted on by gravity follow a `geodesic',
i.e. a path that makes the invariant four-dimensional distance $\int ds$ along the path between fixed initial and final points of the motion extreme.
Free particles that travel at the ultimate speed $c$ also
follow geodesics, are necessarily massless, carry no charge, and cannot spontaneously decay.\footnote{In relativistic quantum theory, no localizable charge can be carried by a massless particle with spin greater than $1/2$, nor can there be a localizable flow of energy and momentum for massless particles with spin greater than $1$. See Steven Weinberg and Edward Witten, ``Limits on massless particles,'' {\it Physics Letters} {\bf B 96 (1-2)}, pp. 59-62 (1980).}

Einstein's General Relativity Theory describes how the classical field $\{g_{\mu \nu}\}$ should vary over space-time.  All `dynamical fields',
 to be consistent with quantum theory, must have corresponding quanta.\footnote{See, for example, Matvey P. Bronstein ``Quantentheorie schwacher Gravitationsfelder'', {\it Physikalische Zeitschrift der Sowjetunion}\,\,{\bf 9},\,pp.\,140-157 (1936). Generally, a dynamical field varies both over space and in time.  Formally, fields which have a kinetic energy term in the Lagrangian for the system
are dynamic.}
We expect that the quantum aspects of gravity will be important near the 
`Planck scale'\;\footnote{Max Planck, ``\"Uber irreversible Strahlungsvorg\"ange.F\"unfte Mitteilung'', {\it K\"oniglich Preussische Akademie der Wissenschaften} (Berlin), Sitzungsberichte,\,pp. 440-480 (1899)%
.}  
$\sqrt{\hbar G/c^3}\approx1.6\times10^{-35}\ m$. Although this is far smaller than the regions we can explore with current accelerators, the very
rarely detected ultra-high
energy cosmic rays may be scattered by this quantum granularity
of space.

\section{Quantum theory}

\subsection{The essence of quantum theory}

Boiled down to its essence, quantum theory follows from a prescription due to Feynman:\footnote{Feynman began thinking of these ideas in 1942.  They are described
in: R.P. Feynman and A.R. Hibbs, {\it Quantum Mechanics and Path Integrals}, McGraw Hill (1965).}

For each particle that was initially observed at $A$ and later observed at $B$,  construct a complex number, called the {\bf transition
amplitude}, as a sum of unimodular complex numbers according to:
\begin{equation}
\left< B | A\right> = N \sum_{paths}\exp{(2\pi i S / h)}\ .
\label{feynman-sum}
\end{equation}
The factor $N$ will be fixed by a `normalization' condition, introduced shortly. Each exponential term in the sum has a phase given by $2\pi S / h$.  The number $h$
is called Planck's constant.  The quantity $S$ is
called the action, defined by a time-integration from $A$ to $B$ of a function $L$:
\begin{equation}
S=\int_A^B L dt \ \ .
\label{eq-action}
\end{equation}

The Feynman sum Eq.\ (\ref{feynman-sum}) is carried over all distinct
paths between $A$ and $B$.\footnote{For an excellent description on how Feynman paths 
are constructed, see Hagen Kleinert, {\it Path Integrals in Quantum Mechanics, Statistics, Polymer Physics, and Financial Markets}, 5th edition, World Scientific (Singapore, 2009).}

The function $L$, called the
Lagrangian,
depends on the particle coordinates and time changes of coordinates for the possible paths between $A$ and $B$.
The Lagrangian is presumed known, and often can be expressed as the particle's kinetic energy
minus its potential energy. Helping to strongly limit the possible Lagrangians is the imposition of the symmetries we observe, such as the Poincar\'{e} symmetry of Relativity.

By reversing the order of the time limits in the
action integral Eq.\,(\ref{eq-action}), the phases of the Feynman amplitudes
change sign, so that time reversal of a transition amplitude is equivalent to 
taking its complex conjugate:
$
\left< B | A\right>=\left< A | B\right>^*$.
Let $B$ range over all possible states into which $A$ may evolve.  Then
$\sum_B\left< A | B\right>\left< B | A\right>$ gives the amplitude for the state $A$
to explore all possible alternatives but then return to itself.  We can take
this amplitude to be unity and thereby fix the magnitude of the normalization constant $N$.
We will then have
\begin{equation}
\sum_B\left< A | B\right>\left< B | A\right>=\sum_B|\left< B | A\right>|^2=1.
\label{eq-norm}
\end{equation}
This relation makes it possible to interpret the magnitude square of the Feynman amplitude as a probability for a given transition.
Doing so creates quantum theory.  
That's it.  All of quantum mechanics follows. 

In contrast to the determinism of Newtonian theory,\footnote{The assumption that systems
have a definite state of existence between interactions would follow
from having only a `single' path dominate the Feynman sum
over paths.} quantum theory gives probabilities for
the result of each measurement of a system. These probabilities
are not simply the result of statistics applied to events.
In quantum theory, a system can be in an interfering combination of
possible realizable events before one of these events is determined
by interactions with another system such as by measurement.\footnote{The fact that certain predictions of quantum theory have intrinsic probabilistic character and that the possible realizable states of
a system retain strange correlations over arbitrarily long distances between particles, greatly disturbed Einstein.  But John von Neumann showed that
quantum theory cannot be trivially subsumed into a bigger deterministic
theory. 
See John von Neumann. 
{\it Mathematical foundations of quantum mechanics}, Princeton
University Press, (1955), Chapter 4.  
For more recent work, see  Roger Colbeck and Renato Renner, ``No extension of quantum theory can have improved predictive power'', 
{\it Nature Communications}\,\,{\bf 2},\,pp.\,411-416 (2011).  
So far, all careful observations are consistent with quantum theory, even
ones that Einstein called 'spooky action at a distance'.}  

If one takes
field quantities as a set of equivalent particle oscillators in each infinitesimal volume of space, with the field amplitudes as the particle displacements, then
quantum field theory follows.

\subsection{The classical limit}

Note that the summation of unit complex numbers with wildly different phases
will tend to cancel (think of adding unit vectors in a plane with
arbitrary angles between them), while a collection of such complex numbers with almost the same
phase tend to add coherently. This observation applied to the Feynman path
sum shows how to take the classical limit, in that those paths causing the least change in the action
$S$ relative to the size of $h$ contribute the most to the probability. 
Classical physics includes only those paths between two
events that minimize $S$.
This is the famous `Principle of Least Action',  from which Newton's laws and Maxwell's
electrodynamics can be derived, after the appropriate choice of $L$.%
\footnote{That non-relativistic quantum mechanics has Newtonian theory as a limit is an example of
the `correspondence limit' which we impose on any new theory in order to
sustain the verified predictions of earlier observations.
After all, Newton's theory predicts natural
processes quite well for massive 
slowly moving bodies, like baseballs, moons, and spacecraft.}  When compared
with quantum theory, Newtonian theory for particles, Maxwell's
electrodynamics, and statistics applied to Newtonian systems
with a large number of particles are together in a realm called `Classical Physics'. 

A {\bf `classical computer'} is a dedicated physical system which transforms
a prepared initial state into a desired output state by applying the equivalent of Boolean logic in one
or more steps between input and output.

\subsection{Superposition}

From the observation that the action satisfies $S_{BA}=S_{AC}+S_{CB}$, it follows from Eq.\,(\ref{feynman-sum}) that
\begin{equation}
\left< B | A\right> = \sum_C\left< B | C\right>\left< C | A\right>\ .
\end{equation}
Quantum amplitudes contain a linear superposition of possible
intermediate states.  If the allowed Feynman paths from $A$ to $B$ are restricted
to only those that pass through two small intermediate regions, say $C_1$
and $C_2$,
there will be interference of the amplitudes constructed to pass through
$C_1$ with those constructed to pass through $C_2$.  This interference can be completely destructive, so that
repeated searches for a particle at $B$ that were launched from $A$ 
come up practically empty.  This effect is observed, and has {\it no}
explanation in classical particle theory.  Yes, you might say, but isn't the
particle a wave?  No, we never observe particles as waves.  
We never find a particle `spread out'.  Rather,
the {\it probability} of finding a particular particle {\it somewhere} can be spread out
over space.  
Individual particles are {\it always} found localized.
Quantum theory lets us calculate these new kinds of probabilities.  New, 
because these probabilities are found by first adding complex amplitudes, a formulation for probabilities unheard of before the second decade of the 1900s.  Addition of amplitudes allows for interference effects, even for a single particle.  This makes the resultant probabilities an intrinsic property of the theory, and not just due to ignorance of states in a more deterministic theory.

\subsection{Wave functions and quantum states}

The Feynman transition amplitude for a particle to leave any earlier location $A$ with
coordinates $x_0$ at time $t_0$
and arrive at $B$ having the location $x$ at time $t$ is called the {\bf wave function} for that particle over the spatial coordinates $x$ at the time $t$:
\begin{equation}
\psi(x,t)=\left<B(x,t) |A(x_0,t_0)\right>\ .
\end{equation}
From  Eq.\,(\ref{eq-norm}), $\int \psi^*(x,t)\psi(x,t)dx=1.$ The symbol
$dx$ in the integral is to be interpreted as the volume element in space.  
We see that $\psi^*(x,t)\psi(x,t)dx$ is the probability of finding
the particle within the volume $dx$. Dirac recognized that 
wave functions may be considered a projection of the `state of the
system' described by a vector denoted $\left|\psi\right>$ onto a specific
state (`eigenstate') of position: $\psi(x,t)=\left<x\right|\psi(t)\left.\right>$.
Each `{\bf quantum state}' $\left|\psi\right>$ can be considered a vector
in a Hilbert space.\footnote{Essentially a vector space with lengths and angles defined, 
but possibly infinite dimensional.} Superposition allows us to
expand the quantum state into a  complete set of basis states: 
\[\left|\psi\right>=\sum_a \left|a\right>\left<a\right|\left.\psi\right>.\] 
(The sum over `$a$' may be given continuous regions as an approximation
to discrete sums which are dense in those regions.)

From the Feynman path sums, the state of a system evolves in time according
to a linear transformation
\begin{equation}
\left|\psi(t)\right>=U(t,t_0)\left|\psi(t_0)\right> \ ,
\end{equation}
where, to keep the total probability of finding the particle anywhere unity,
the operation $U$ must be unitary: $U^{\dagger}U=1$.  
(The `dagger' here performs a transpose-complex-conjugate operation, 
rather than just complex-conjugation, to include
cases in which $\psi$ is taken to have components.)
The Feynman path summation divided into small time steps means we can write $U=\exp(-i\int H dt/\hbar)$.  
(The sign in the exponent
is conventional. The constant $\hbar=h/(2\pi)$.)
The operator $H$, called the Hamiltonian, satisfies the `Hermiticity condition' $H=H^{\dagger}$.
In the language of Lie groups, $H$ is a generator of
time translations.
For small shifts in time,  $\psi$ satisfies
a linear equation:
\begin{equation}
i \hbar \partial_t \psi(x,t) = H\psi(x,t) \ .
\end{equation}
This is a wave equation, which formed the basis of the dynamics of 
quantum theory originated by Schr\"{o}dinger.  


\subsection{Particles in relativistic quantum theory}

Our present quantum theory incorporates Einstein's Special Theory of Relativity.\footnote{A. Einstein, ``Zur Elektrodynamik bewegter K{\"o}rper'', {\it Annalen der Physik},\,{\bf 17},\,pp.\,891-921 (1905).} 
P.A.M. Dirac, recognizing that
Relativity requires that physical laws be expressible with space and time on an equal footing, wrote
the Hamiltonian as a linear operator in the generators of space translation,
so that the wave equation took the form\footnote{P.A.M. Dirac, ``A Quantum Theory of Electrons'', Part I \& II, {\it Proc.Roy.Soc. (London)},\,{\bf A\,17} \& {\bf A\,118},\,pp.\,610-624 \&\,pp.\,351-361 (1928).} 
\begin{equation}
\sum_{\mu=0}^3\gamma^{\mu}(i\hbar \partial_{\mu}-(e/c)A_{\mu})\psi=mc\psi \ .
\label{dirac-eq}
\end{equation}
When the fields $A_{\mu}$ vanish, there are plane wave solutions 
$\psi\propto \exp{(-ip_{\mu}x^{\mu})/\hbar)}$, so that
$g_{\mu\nu}i\hbar \partial^{\mu}i\hbar \partial^{\nu}\psi=
{p}_{\mu}{p}^{\mu}\psi
=m^2c^2\psi$, and 
the $\gamma$'s must satisfy
\[{\gamma^{\mu}\gamma^{\nu}+\gamma^{\nu}\gamma^{\mu}=2g^{\mu \nu}I}.\]
If we add the assumption of reflection symmetry, the $\gamma$'s are square matrices with even dimension at least four.
Taking the $\gamma$'s to be dimension
four, and  the fields $A_{\mu}$ as the electromagnetic vector potentials due to
other charges, the Dirac equation
very accurately describes electrons in the field of other charges, and therefore atomic structure and, in principle,
all of chemistry and molecular biology.  The components of the electron wave function
can be decomposed into two pairs, each pair corresponding to
the two possible intrinsic spin directions measurable,
and the combined pair corresponding to the electron carrying positive
or negative energy.
As an indication of the profound reach gained by merging quantum theory and Relativity, 
Dirac was
able to show that the electron spin and its magnetic moment followed from relativistic quantum theory, and that antimatter must exist, a prediction before anyone dreamed of the concept.

The possibility that fundamental particles can be created and destroyed
is included into quantum theory by taking the particle wave functions
and interacting fields
as quantum fields,  entering into the action $S$ with their own dynamics.
We find that if disturbed, particle pairs can even `bubble' out of empty space.  
The time-and-space-reversed wave function for a particle describes the forward progression of a corresponding antiparticle.  This becomes the `CPT Theorem' in quantum field theory, referring to the operations of charge conjugation, parity transformation,
and time reversal.

Quantum field theory distinguishes particles with half-odd
integer spin, called `fermions', from those with integer spin, called `bosons'.%
\footnote{Particles must have
quantized spin with length $\sqrt{s(s+1)}\hbar$ and projection along some measurement axis of
$\mu\hbar$, where $s$ is either a half or whole integer, and $-s\le\mu\le s$.  It is conventional to
use the label $s$ to characterize the particle spin, as in ``The electron has spin $1/2$''.  
Particles that move
at the speed of $c$ have only two projections of
their spin, called their `helicities', either along their momentum,
or in the opposite direction. The characteristic properties of particles following from relativistic quantum theory were first
described by Eugene Wigner in ``On Unitary Representations of the Inhomogeneous Lorentz Group'',  {\it Ann. Math.}\,{\bf 40 (1)},\,pp.\,149-204 (1939). }
The quantum field in a three-dimensional space and
associated with a pair of identical particles may undergo a phase
change when those
two particles are exchanged: $\left|\psi(1,2)\right>=(-1)^{2s}\left|\psi(2,1)\right>$.  If the particles are fermions ($s=1/2,3/2,\cdots$)
the phase change is $-1$, while no phase change occurs
if the particles are bosons ($s=0,1,2,\cdots$).  This means no two fermions of the same type
(such as electrons in atoms) can occupy 
the same quantum state.  This is the `Pauli Exclusion Principle'.  Any number 
of bosons of the same type can be in the same quantum state (e.g. photons in lasers).

The fundamental particles making up 
the structure of materials currently appear to be three generations
of the doublet electron-neutrino,\footnote{Our observations
of the sky together with General Relativistic cosmology seem not to allow more than three generations.} and three generations of a doublet of quarks, all fermions. The family of electrons and neutrinos are called
`leptons'. Each generation of quarks comes in one of three 
distinct varieties according to their `color charge'. The bound state of a `red', `green', and `blue' quark and any other `color-neutral' combination of an odd number of quarks generates  a `baryon', such as the familiar
proton and neutron. A zoo of more fleeting particles exist, including `mesons' coming from bound color-neutral quark-anitiquark systems. The large family of baryons and mesons, all strongly interacting
particles, are called `hadrons'. In the Standard Model, leptons have
no direct strong interactions.

\subsection{Interactions in quantum theory}

All the observed interactions of one particle with another can be categorized by the so-called
strong, electromagnetic, weak, and gravitational forces.\footnote{The electromagnetic and weak interactions
were linked, principally by the work of Salam, Glashow, Weinberg, Higgs, 
't\,Hooft, and Veltman, from 1964 to 1975.}

The numerical strength of a particle's interactions with other particles is always associated with an intrinsic property called its `charge'.   For each category of
interaction,  
there is one or more corresponding charges.  If
the total charge of a closed physical system is preserved 
during a sequence of interactions within that system,
we say the charge has been `conserved'. In nature, all charges are quantized, i.e.,  they come from a countable set.\footnote{Dirac showed that if magnetic monopoles exist, then electric charge must be quantized.  See  Paul Dirac, ``Quantised Singularities in the Electromagnetic Field'', {\it Proc. Roy. Soc. (London)}\,{\bf A\,133},\,pp.\,60-71 (1931).}

The existence of conserved and localizable charges 
means one can always define an {\bf interaction field} that has those charges as its source, using the following argument:
If $\{j^{\mu}\}=\{\rho c,\rho\vec{v}\}$ represents the charge density
and current density for a set of charges, then
the local conservation of the total charge, $Q\equiv \int \rho d^3x$,
can be read from $\partial_{\mu}j^{\mu}=0$.
But this 
implies the existence of
an {\it `interaction field'} 
$\{F^{\mu\nu}\}$, 
antisymmetric in its indices, satisfying
$\partial_{\kappa}F^{\kappa\nu}\propto j^{\nu}$.  An associated field,
${F^*}_{\mu\nu}\equiv(1/2)\epsilon_{\mu\nu\kappa\lambda}F^{\kappa\lambda}$
defines a `dual' conserved charge with current $j^{*\nu}\propto \partial_{\mu}F^{*\mu\nu}$.\footnote{
The $\{\epsilon_{\mu\nu\kappa\lambda}\}$ is the
completely antisymmetric tensor in four dimensions, with $\epsilon_{0123}=1$, 
called the Levi-Civita symbol.  Like ${g_{\mu\nu}}$, its components are invariant
under a proper Poincar\'e transformation.}
If no such dual charge exists in a region of space, then the field $\{F^{\mu\nu}\}$, assumed to carry no intrinsic charge itself,\footnote{
For the strong-interaction charges, the field $\{F^{\mu\nu}\}$ does carry charge.  It then can interact with itself, making the wave
equation for dynamical field theory constructed for $\{F^{\mu\nu}\}$
necessarily non-linear.}
can be expressed in terms of a vector field $\{A^{\mu}\}$ by $F^{\mu\nu}=\partial^{\mu}A^{\nu}-\partial^{\nu}A^{\mu}$.
The field $\{A^{\mu}\}$ is called the {\bf `gauge field'} 
going with the corresponding charge. 
Gauge fields are not uniquely determined, but
may be transformed into new fields $\{{A'}^{\mu}\}$ which have the same interaction field $\{F^{\mu\nu}\}$ by adding a gradient: ${A'}^{\mu}=A^{\mu}+\partial^{\mu}\Lambda$.
The choice of the `gauge function' $\Lambda(x)$ is open, 
provided $\oint \partial_{\mu}\Lambda dx^{\mu}$ vanishes
for all closed loops in regions where the
gauge field acts.  
Theories whose predictions
are independent of the choice of gauge 
have {\bf `gauge symmetry'}.\footnote{The use of `gauge symmetry' 
was introduced by Herman Weyl in his consideration
of theories with invariance in the scale of length. (H. Weyl, ``Gravitation und Elektrizitat'', {\it Sitzungsber. Preuss. Akad.
Wiss.},\,pp.\,465-480 (1918).)}

Conventional theory describes
particle interactions by introducing interaction fields which `mediate'
the effect of one charge on another. 
We say each particle with a charge of some kind `creates' an interaction field in the space around it, and
that field acts on other particles having the same kind of charge. 
In the case of electromagnetic interactions, the interaction field is $\{F^{\mu\nu}\}$
with components that are  the electric and magnetic field, while
the gauge field $\{A^{\mu}\}$ is called the electromagnetic vector potential.
Maxwell's equations, $\partial_{\kappa}F^{\nu\kappa}=(4\pi/ c) j^{\nu}$
and $\partial_{\kappa}F^{*\nu\kappa}=0$, then express two conditions: Electric charge is conserved locally, and there is no
observable local magnetic charge.
How particles react to other charges requires knowledge of the
dynamics for those particles. Dynamics is incorporated into quantum theory.

In quantum theory, a second kind of gauge transformation occurs
when the phase of particle wave functions are shifted.  A constant
shift has no observable effect.  But making a shift in phase
which depends on location will introduce a relative phase 
between wave components.  If those component waves
converge, their interference is observable in the associated
particle probability.  Now, if, along with the phase shift,
a shift in the derivatives of the wave function occurs, one
can make the combined shifts cancel.  This is the
property built into {\bf `gauge symmetric quantum theories'}. In fact,
all the interactions among fundamental particles have been found to
follow from theories which satisfy gauge symmetry! 

Another property of our current dynamical theories can be called the {\bf principle of quasi-local interactions}: The 
known interactions of 
one particle with another
can be described by `quasi-local' effects, castable into a form that requires only knowledge of the
fields of other particles in a small local space-time neighborhood of the affected particle. These fields are the gauge fields described above.
Consider the free-electron Dirac equation $\hbar \gamma^{\mu}i\partial_{\mu}\psi=mc\psi$.  A gauge
transformation of the second kind on the wave function
can be expressed as $\psi'(x)=\exp{\left[-i(e/(\hbar c))\Lambda(x)\right]}\psi(x).$
The free-particle wave equation becomes $\gamma^{\mu}(i\hbar\partial_{\mu}-(e/c)\partial_{\mu}\Lambda)\psi'=mc\psi'.$  Gauge
symmetry can be enforced by adding to the derivative term a gauge
field $A_{\mu}$ which undergoes a gauge transformation
of the first kind: ${A'}^{\mu}=A^{\mu}+\partial^{\mu} \Lambda$.  We arrive at
the full Dirac equation (\ref{dirac-eq}).
This technique for introducing interactions is referred to as the `minimal coupling principle'.\footnote{  
Gauge symmetry in quantum theory can be re-expressed
in terms of the action of `covariant' derivatives $D_{\mu}=\partial_{\mu}+i(e/(\hbar c))A_{\mu}$, acting on a quantum state for a particle.  In this interpretation,  the interactions arise from the behavior of quantum states by parallel transport across space. When the gauge fields themselves are taken to be operators on the internal components of a quantum state, the gauge group elements may not commute.  These kind of `non-abelian' gauge fields were introduced by C.N. Yang and R. Mills (``Conservation of Isotopic Spin and Isotopic Gauge Invariance'', {\it Phys.Rev.}\,{\bf 96 (1)},\,pp.\,191-195 (1954)) and are used in the Standard Model to describe interactions between fundamental particles grouped into families.  For example, the quark color charge follows from an $SU_3$ gauge symmetry.}  

A marvelous theorem was derived by Emmy Noether,\footnote{E. Noether, ``Invariante Variationsprobleme'', {\it Nachr. D. K\"onig. Gesellsch. D. Wiss. Zu G\"ottingen, Math-phys. Klasse},\,pp.\,235-257 (1918).} who showed that symmetries of our theories based on continuous groups of transformations,
such as the Poincar\'e group and the gauge transformations, lead to
conservation laws.  In the case of the Poincar\'e symmetry, the conserved quantities are total energy-momentum, total angular momentum, and the velocity of the center-of-energy.  An important example is the symmetry under time translation: If experiments done now with a given system have the same set of results as those done at any time later, then the system's energy is fixed.  Symmetry under constant phase shifts
of a lepton or baryon wave function makes lepton and baryon number conservation.
Gauge symmetry makes the corresponding charges conserved.

As the gauge fields have their own dynamics, quantum theory requires that
gauge fields be quantized.  That means that the interactions between
material particles occur only
by the exchange of quanta.  These quanta are necessarily bosons.
For electromagnetic interactions, the gauge-field quantum is the photon.  
The photon at present appears to travel at the maximum speed in Relativity, has unit spin, and carries no electric charge.\footnote{A particle with charge will carry energy
associated with the field of that charge, and therefore, if it can be separated from other particles, must have mass, and must move slower than the universal speed c.}  For the strong interactions between quarks, the quanta of the field are called gluons.  Gluons also have unit spin but they carry various `color' charges.   By having charge, gluons can directly interact
with themselves, making their dynamics more complicated than for photons.  For example,  
the gluon fields, through their self-interaction, can form flux tubes between quarks.

\subsection{Prepared states and measurement}

Each possible quantum state of
a system is referred to as a `pure quantum state', as contrasted to a `mixed quantum state', for which we may only know probabilities for the system to be
in given quantum states.  A pure quantum state made from a superposition of component states is called a `coherent quantum state' when all the phases between its various component states are known to be fixed.

A quantum system is prepared by first selecting a physical system, isolating the system from unwanted interactions, determining its initial configuration, and then stimulating or allowing the system to approach a desired initial state.

Isolating a system and determining its initial configuration are often daunting tasks.
The state of most macrosystems will be practically impossible to completely
specify.  Some interactions,
such as those from stray fields or background radiation, may be
difficult or impossible to eliminate.  In addition, various possible components
within an
isolated system can transform and evolve even when isolated.  However, fundamental
particles will, by definition, be stable, at least over times much longer than observational periods.  Also, some bound systems of fundamental particles will be quasi-stable if the energy need to excite the system is large compared  to the energies available.
After isolation, a system will evolve by
quantum dynamics following a unitary transformation, and may eventually become a 
`steady state', i.e. one with no change in probability densities
for its particles, if these were observed.

Consider the expansion of a pure quantum state into component states which together span
the system's Hilbert space:
\begin{equation}
\left|\psi\right>=\sum_i\alpha_i \left|\phi_i\right>.
\label{coher-state}
\end{equation}
If the phases
between two or more components of the quantum state are related, then these components are 
said to be in `{\bf coherence}'.  Quantum interference between various possible outcomes of a measurement requires some coherence in a quantum state. 
The states $\left|\phi_1\right>$ and $\left|\phi_2\right>$ might be two possible interfering states of a single electron, or even a
trapped atom.  The two states of the atom might have opposite motions, so that the wave function for each state can oscillate back and forth across the trap. Then
the probability distribution of finding the atom at a specific location within the trap shows an
interference pattern.\footnote{This game was played using a Beryllium atom by Dr. Christopher Monroe and colleagues at the National Institute of Standards and Technology, Boulder, Colorado. See C. Monroe {\it et al.}, ``A `Schr\"odinger Cat' Superposition of an Atom,'' {\it Science}\,\,{\bf 272}, pp.\,1131-1136 (May 24, 1996).  Some members of the press mis-represented the observation as indicating that one atom can be found in two places at once.  For example, see Malcolm W. Browne's article ``Physicists Put Atom In 2 Places At Once'', published in the New York Times, May 28, 1996.}  This quantum effect, however, is no different in principle than that seen as an interference pattern made by the bright spots of light on the
surface of a phosphor plate, those spots produced by electrons passing through
two slits in a screen, one electron at a time, and then hitting the phosphor.

Starting with a set of identically prepared systems in a coherent state represented by Eq.\,(\ref{coher-state}), measurements of the observable 
$A$ will have an average $\sum_{ij}\alpha_i^*\alpha_j \left<\phi_i\right|A\left|\phi_j\right>$. Interference will arise from terms for which $\left<\phi_i\right|A\left|\phi_j\right>$ are not zero for $i\ne j$.
However, if a quantum system interacts with another system or
with a measuring device, some or all of the components of residual quantum states for that system
may be left with no well-defined phase relationships.  This is a process of {\bf `decoherence'}.
During the measurement, information is transferred between the system
and the measuring device, and some may be lost to the environment.

One of the important measurements of a system locates the position of
particles.  After a number of such measurements in each small
region $dx$ of space, we find a distribution of positions.  For one
particle, the wave function determines the probability density for position across space, so the distribution of measured positions is predicted to be an approximation of $\psi(x)^*\psi(x) dx$.  The average position over all
space is predicted to be $\left<\psi\right|x\left|\psi\right>$.
More generally, each distinct measurement of a property of the system can be associated
with a Hermitian operator $A$ that acts on wave functions for the system $\psi$
as follows:
The average value of $A$
will be 
\begin{equation}
\left<A\right>=\int \psi^{\dagger}(x,t)A(x,\hat{p}_x)\psi(x,t)dx\ ,
\end{equation}
wherein $\hat{p}_x$ is taken proportional to the space translation operator
in accord with Noether's Theorem.\footnote{The proportionality
constant is fixed by noting that if a free particle is left unobserved, then
within some bounded region its wave function becomes a `plane' wave $\phi\propto \exp{((i{p_x}{x}-iEt)/\hbar})$, for which $\hat{p}_x\phi=-i\hbar\partial_x\phi$. The order of non-commuting operators in $A$ must be determined by physical arguments.}
The operators $A$ may also act on the spin and 
other components of the wave function.  

Those states $\left|a\right>$ satisfying
\[A\left|a\right>=a\left|a\right>\] are called the `eigenstates' of $A$ and 
$a$ an `eigenvalue'.  For Hermitian operators $A$, i.e. $A^{\dagger}=A$,
the eigenvalues $a$ will be real numbers, and therefore
each is a value which may result from a measurement. The elements of the set 
$\{\left|a\right>\}$
for distinct values $a$ will be `orthogonal', i.e. $\left<a'|a\right>=\delta_{a'a}$, and `complete', i.e. they span the space of possible states, expressible as
$\sum_a  \left|a\right> \left<a\right|=I$ by reading from $\left|\psi\right>=\sum_a  
\left|a\right> \left<a\right.\left|\psi\right>.$
The measured values of $A$ will have an uncertainty defined by
$
\Delta A\equiv \sqrt{\left< (A-\left<A\right>)^2\right>} \ .
$
This means that after measurement of $A$ for a large number of identically prepared systems, the observed values will be distributed 
around the average with a `width' of $\Delta A$.

After a single measurement of the observable $A$ for a system in a pure quantum state $\left|\psi\right>$ that has $A$ as one of its observables, one of the eigenvalues of $A$, say $a$, will be found, and the system will be left in
the state $\left|a\right>$.  The effect of measurement can be represented by a {\bf `projection operation'}: $P_a\equiv\left|a\right>\left<a\right|$.  The measurement of $A$ has `collapsed' the
quantum state to $\left|a\right>\propto P_a\left|\psi\right>$. 
The
collapse evidently does not preserve unitarity for the system, expressed by $\left|\psi(t)\right>=U(t,t_0)\left|\psi(t_0)\right>$, unless the system was already in an eigenstate of $A$.  Quantum unitarity applies to isolated systems. The measurement
process has involved another interacting system which reduced
the system's available states in a subsequent measurement.  

The interaction (called the `coupling') between two systems during a measurement may cause one or more of the phase 
differences between the components in the
resulting quantum states to become indeterminate, especially
likely if the measuring device is macroscopic.  After
the measurement,  the system may be left in a `mixed' state, for which only the probabilities $p_k$ for any particular pure quantum state $\left|\psi_k\right>$ are known.  Then a subsequent measurement of the observable $A$ will have an average value 
of $\left<A\right>=\sum_k p_k \left<\psi_k\right| A \left|\psi_k\right>$. This expression can be usefully re-written in terms
of a `density operator', defined by $\hat{\rho}\equiv\sum_k p_k \left|\psi_k\right> \left<\psi_k\right|$, so that $\left<A\right>=Tr{(\hat{\rho}A)}.$ In this way, the choice of the mixed state is left implicit.  A pure state can
then be simply characterized by $\hat{\rho}^2=\hat{\rho}$.

\subsection{Entangled states}

An {\bf `entangled quantum system'}, by definition,
has two or more particles in a quantum state which cannot
be factorized into states for each particle.\footnote{There is a special caution for quantum states describing photons, in
that the number of photons is not fixed, but rather has an uncertainty which
increases as the phase of the electromagnetic wave becomes more definite.}  For
example, if we let the quantum state $\left |\mu_{\kappa}\right>$ represent
the electron labeled by $\kappa$ and having a spin projection along the $z$-axis of $\left(\mu-1/2\right)\hbar$,
then one of the possible entangled two-electron states can be written
\[\left |\psi_0\right>=(1/\sqrt{2})(\left |0_1\right>\left |1_2\right>-\left |1_1\right>\left |0_2\right>),\]  
which happens to have a total spin of zero.

The outcome of a measurement
of one of the electrons in an entangled pair will be correlated with the outcome of measurement of the other, even when they are far apart.  This kind of correlation also occurs
under classical conditions. Suppose you put a jack in one envelope and a queen in another.  Now send one of the envelopes to one friend, the second envelope to a second friend.  If one friend opens your envelope and finds a jack, then your other friend must find a queen even before
hearing from your first friend.  However,
there is a twist in the quantum world.  Take the case when a pair of electrons
is prepared in a zero total spin state along a $z$-axis expressed above, and then the electrons are allowed to move far apart.  Next, while the electrons are in flight,
have one of the distant observers rotate her electron-spin measuring apparatus away from the $z$-axis direction to an angle of her choosing, i.e. the
first distant observer makes a {\bf `delayed-choice experiment'}.\footnote{
If the decision on how a component of a system is measured
comes after that system has had sufficient time to cause
interference between quantum alternatives for that component, then
this becomes a delayed-choice experiment as introduced by
John Wheeler in {\it Mathematical Foundations of Quantum Theory}, edited by A.R. Marlow, Academic Press (1978).}  If this first observer finds an electron aligned along her new axis,  then the second observer, far away, will find the other electron aligned along the negative direction of the NEW axis constructed
by the first observer.  Now
we, on first hearing and with our classical thinking, SHOULD be surprised!  Even so, this is the
way nature acts.
The result does NOT mean that the pair interacted after traveling apart, nor was there `superluminal' transmission of information.\footnote{Information transmitted by a wave disturbance that started at a certain time cannot be transferred faster than the outgoing wavefront from that disturbance.  In Special Relativity, the speed of the wavefront, also called the signal speed, is always less than or equal to the universal limiting
speed, $c$.  There is no such restriction on the group velocity or the phase velocity of the wave. }  
This suggestion of faster-than-light signaling is a misinterpretation of quantum theory, and such information transfer has not been seen.  Rather, those who say so are likely to have been tripped up by picturing each unobserved electron as being localized  between observations!

\subsection{Non-classical interactions}

There are interactions predicted by quantum theory without classical explanation.
Yakir Aharonov and David Bohm\footnote{Y. Aharonov and D. Bohm, ``Significance of Electromagnetic Potentials in the Quantum Theory'', {\it Phys.Rev.}\,{\bf 115},\,pp.\,485-491 (1959).} showed that a single electron
wave which never enters a region of electric or magnetic field could never-the-less have a measurable shift in the probability of finding that electron after an electric or magnetic field changes in the excluded region.  The effect occurs, for example, when the electron passes on either side but does not enter a tube where a magnetic field is confined.  The difference in phase of that wave when followed around a closed loop is given by $e\oint A^{\mu}dx_{\mu}/(\hbar c)$, where $\{A^{\mu}\}$ is the electromagnetic
potential. This is the flux of magnetic field somewhere inside the loop.\footnote{Mandelstam re-expressed the local interaction with $\{A^{\mu}\}$ as a non-local effect of the electric and magnetic fields, i.e.  a topological effect of fields over space-time. See Stanley Mandelstam, ``Quantum electrodynamics without potentials'', {\it Ann. Phys.}\,{\bf 19},\,pp.\,1-24 (1962).} The shift in the observed interference pattern produced by the electrons when the magnetic flux is turned on has no explanation in classical physics.

Measurement of a system may disturb the system.  If the measurement process
transfers complete information about a system, that system will
no longer contain entangled states.  This effect leads
to the `no-cloning theorem',\footnote{
Wojciech Zurek, ``A Single Quantum Cannot be Cloned'', {\it Nature}\,{\bf 299},\,pp.\,802-803 (1982); Dennis Dieks, ``Communication by EPR devices'', {\it Physics Letters}\,{\bf A\,92\,(6)},\,pp.\,271-272 (1982).}
the statement that a general quantum system containing some coherence cannot be identically copied.  If a copy of a quantum state could be made, then we could defeat the interfering effect of measurement by first making a copy, and then measuring the copy, leaving the original system undisturbed.

The wave function for a particle confined to a fixed region of space and initially localized to a much smaller part of that region and then left with no 
external interaction will
diffuse outward in space as time progresses.  The wave for an unobserved particle will spread over the entire allowed region, and
eventually the probability for finding the particle in any small location
will have no measurable change in time, and its quantum wave function
will be steady.\footnote{Steady wave functions necessarily
have a sinusoidal time dependence through a factor of the form $\exp{(-i\omega t)}$, making
the probability $\psi^{\dagger}\psi dx$ time independent.}

A localized and isolated physical system will have
denumerable (`quantized') possible values for its measurable energies
and momenta.
Periodicities of the wave function also enforce quantization
if there is a closed path over which the corresponding particle
can move.  For example, periodicity in the azimuthal angle in the wave function makes the measured values of the projection of the orbital angular momentum along a measurement axis denumerable.  

Suppose two observables $A$ and $B$ for a given system in the
state $\left|\psi\right>$
are measured in a certain time order.
If these two measurements are repeated for identically prepared systems, a change in the
order of measurement may change the probability for finding a given value
for the second observable.  In general, one can show that
the uncertainties satisfy $\Delta A\ \Delta B\ge(1/2)\left|\left<AB-BA\right>\right|$. This is called the uncertainty
principle of Heisenberg.  If the `commutator' $[A,B]\equiv (AB-BA)$ vanishes, then
the observables $A$ and $B$ may be measured `simultaneously', i.e.
without the measurement of one affecting the results of measuring the
second.  The state of a physical system can be labeled by a 
set of measured values for a maximal set of mutually commuting 
observables that are also conserved over time.

\subsection{Quantum theory for complex systems}

After a relaxation time for a system containing a large number of interacting particles, the most likely distribution of particles
in the available quantized energy states will be those
that tend to maximize, under the physical constraints, the multiplicity $W$, simply because
as the system evolves through various configurations, it will spend most of
its time in those configurations which have many ways of being constructed.  
One can then show the following:\footnote{Ludwig Boltzmann in ``\"Uber die beziehung dem zweiten Haubtsatze der mechanischen W\"armetheorie und der Wahrscheinlichkeitsrechnung respektive den S\"atzen \"uber das W\"armegleichgewicht'', {\it Sitzungsberichte der Akademie der Wissenschaften zu Wien}\,{\bf 76}, pp.\,373-435 (1877).} Divide the system
into a large number of subsystems, labeled by $i$, which have possible energy states equally likely to accept energy from its neighbors.  
(One such choice of subsystem could be the identifiable particles in the system.)
Suppose the number $g_i$ of all the possible states of any one of these subsystems which have energy near  $\epsilon_i$ is much larger than the actual the number of subsystems $n_i$ carrying those nearby energies, and that $n_i$ itself is large. Then near thermodynamic equilibrium the number of subsystems with energy near $\epsilon_i$
is given by $n_i\propto g_i\exp{(-\epsilon_i/(kT))}$, where $T$ is the temperature of the system.

Interactions from the outside can change the
total energy, $E=\sum_i n_i \epsilon_i$, of a system 
either by changing the `occupation numbers' $\{n_i\}$, and/or by changing
the energies $\{\epsilon_i\}$ of the quantum states.  The first kind of change is
heat transfer and the second is work transfer.
By increasing
the multiplicity of the system, putting heat into a system is a `disordering process'.  Work involves changing the particle energies by
changing the volume of the system, without moving particles between quantum states.\footnote{If particles remain in their quantum states, no heat is transferred, and the process is called `adiabatic'.}
These ideas incorporate the first and second law of thermodynamics.  

In terms of information, the second law of thermodynamics implies that {\it if two systems interact, each with fixed volumes, then that
system of the two which has the smaller variation in its information content as its total energy changes will tend to spontaneously transfer information into the second system.}\footnote{If a system
near thermal equilibrium is held at fixed volume and a small amount of energy $dE$ is put in, causing an increase in its information content by $dI$, then the ratio $dE/dI$ turns out to be proportional to the temperature of that system. The spontaneous flow of information, i.e. non-forced flow, results from statistical likelihood.}

These
are important concepts for quantum computers, as there is an intimate connection between entropy,  information, decoherence, wave function collapse, and heat from memory loss.

\section{Quantum computation}

Feynman\footnote{Richard Feynman, ``Simulating Physics with Computers'', 
{\it International Journal of Theoretical Physics}\,{\bf 21},\,pp.\,467-488 (1982).}
considered the possibility that we might take advantage
of quantum systems to perform computations quicker than
so-called classical computers. Modern classical 
computers
use bistable systems to store information, and
logical gates to perform Boolean
operations on sets of ones and zeros.%
\footnote{These discrete-level computers are often referred to as `digital', in contrast to `analog' computers that
use internal signals that are assumed to vary smoothly with time. Mechanical computers, which work by the movement and interaction of shaped
objects, and molecular computers, that work through molecular interactions and transformations, are a mixed breed.  The
phrase 
digital computer, referring to counts base ten, can now mean any device which manipulates information by discrete changes.  These days, the changes
are made in systems which can flip between {\it off} and {\it on} in a specified
clock time, i.e. a binary
coding. By using such switching to encode information, digital computers
can be more tolerant of a small amount of noise than analog devices.
Shannon and Hartley
showed 
that the maximum number of bits per second that can be transmitted
from one storage location to another is given by $B\log_2{(1+S/N)}$, where
$B$ is the bandwidth (in cycles per second), $S$ is the average signal power, and $N$ is the average noise power.  See R.V. L. Hartley, ``Transmission of Information'', {\it Bell System Technical Journal} (July 1928); C.E. Shannon,``Communication in the presence of noise'', {\it Proc. Institute of Radio Engineers} \,{\bf 37 (1)},\,pp.10-21 (January 1949).}
For some
problems involving numbers with $n$ digits and that may require
solution times that rise exponentially with $n$ when
performed on computers using only Boolean logic, the
computation on a quantum computer may take
times that rise no faster than a power of $n$. 
Below are some of the special consequences of quantum theory for
quantum computers and communications:

The simplest system for the storage of information
gives only two possible values by a measurement.  These
values can be taken as $0$ or $1$, in which case
the states are called $\left|0\right>$ and $\left|1\right>$.  
Classically, such a system stores one
bit of information.  A quantum system can be constructed
that has only these two values for the outcome
of a measurement, but whose quantum state is
a linear combination of the two possible outcome states 
$\left|0\right>$ and $\left|1\right>$:
\[\left|q\right>=\alpha\left|0\right>+\beta\left|1\right>.\]
This state is called a `qubit', where $\alpha$ and $\beta$ are complex numbers satisfying $\left|\alpha\right|^2+\left|\beta\right|^2=1$.  An alternative
parameterization takes $\alpha=\cos{(\theta/2)}$ and 
$\beta=\exp{(i\phi)}\sin{(\theta/2)}$. Evidently, the possible qubit
states can be pictured as points on a unit sphere (called the 
`Bloch sphere') with $\left|0\right>$ at the north pole and $\left|1\right>$ 
at the south.
Two-valued qubit states are
easily realized in nature: The electron spin has only two
possible projection values $\pm\ 1/2 \hbar$, and the photon
has only two possible helicity values $\pm\ 1 \hbar$. 

As it is always possible to expand an arbitrary
quantum state into a basis set for that state's Hilbert space, 
$N$-particle states in a quantum computer
can be made by constructing these quantum states from a linear
combination of the states for each of the $N$ particles.  Taking these
particles to have only two internal quantum states, the state of 
the computer is expressible by
\begin{align}
\left|\psi_N\right>&=\sum_{\{i_k=0,1\}}\beta_{i_1i_2i_3,\cdots i_N}
\left|{i_1}\right>_1\left|{i_2}\right>_2\left|{i_3}\right>_3\cdots\left|{i_N}\right>_N \nonumber \\
&=\sum_{i=1}^{2^N}\beta_{i}
\left|{i_1}{i_2}{i_3}\cdots{i_N}\right> \ \text{with}\ \sum_{i}\left|\beta_i\right|^2=1 \ .
\label{N-state}
\end{align}
In the second line of the equation, the product base state is represented  
in a shortened form, in which the order of the $0$'s and $1$'s corresponds to the labeling
of each of the separate qubits, and $i={i_1}{i_2}{i_3}\cdots{i_N}$
is a binary number constructed from the $i$'s. If the quantum state $\left|\psi_N\right>$ cannot be factorized, it harbors entanglement. Quantum computation takes advantage of entanglement
within those states.\footnote{See, for example, Richard Jozsa and Noah Linden, ``On the role of entanglement in quantum computational speed-up'', {\it Proceedings of the Royal Society A: Mathematical, Physical and Engineering Sciences} {\bf 459}\,(2036),\,pp.\,2011-2032 (2002).}   It follows that a useful initial state of a quantum computer has at least a subset of particles prepared in one of the {\bf maximally entangled states}, i.e. states with equal probability for all possible configurations of its component particles, making it also that state which has maximum information content.\footnote{One learns most when the outcomes are least predictable!} The maximally entangled states made from qubits as in Eq. (\ref{N-state}) will have all $\left|\beta_i\right|^2=1/N$, leaving $(2^N-1)$ free relative phases between the basis states.\footnote{These maximally entanglement states are also called `generalized Bell states'.} 

To sustain coherence, quantum computers must 
operate on
the input information
stored in quantum states by unitary transformations.  
In the following, the substage of a quantum computation holding 
the intermediate state of a calculation will be called $\left|\psi(k)\right>$,
where $k$ labels a particular intermediate state, with $k=0$ labeling
the initial state.
For a given quantum computer, a solution to a solvable problem
is a unitary transformation $U_S$ that carries the input quantum state
 $\left|\psi(0)\right>$
encoding the required initial data into an output quantum state
that carries the information about the solution, at least in probabilistic terms.
To be a non-classical computer, at least some the intermediate states must be entangled. 
It is possible that  $U_S$   
can be decomposed
into a finite product of simpler or more universal unitary operations: $U_S=\prod_i U_{g_i}$,
where the set $\{U_{g_i}\}$ are called `quantum gates', a generalization of
classical logic gates.  Each term in the product acts on the state $\left|\psi(k)\right>$ 
left by the previous
operation labeled by $k$ and produces $\left|\psi(k+1)\right>$.  

Since
a general unitary transformation will contain continuous parameters,
$U_S$ might only be approximated  by a finite sequence of quantum
gates. In the classical case, all Boolean operations on a set of bits can be
performed by a combination of NAND gates.  This makes
NAND gates universal for classical computing.  The same is true of NOR gates.
In the quantum case, there are `universal' sets of simple gates
that can be used to build arbitrarily close representations of
a general unitary transformation, such as $U_S$.  (Arbitrarily close here means that if $V_S$ is the approximation, then 
$\left|\left<\psi\right|(U_S-V_S)\left|\psi\right>\right|^2$ 
is a number that
can be made arbitrarily small for all $\left|\psi\right>$ by
increasing the number of universal gates used in $V_S$.)

Quantum gates acting on a single qubit can all be represented by a general unitary transformation $U_{\vec{\theta}}$ which is
an arbitrary rotation in Hilbert space: \[U_{\vec{\theta}}=\exp{(i\theta\ \hat{n}\cdot \vec{\sigma}/2)}=\cos{(\theta/2)}\ I+i\hat{n}\cdot \vec{\sigma}\sin{(\theta/2)},\]
where $\theta$ is an angle of rotation around an axis fixed by
the direction $\hat{n}$,  and
the $\{\sigma_i\ ;\ i=1,2,3\}$ are the Pauli matrices,
\begin{align*}
\sigma_x=\left[
\begin{array}{cc}
0 &1 \\
1 & 0
\end{array}
\right] \ \
&\ , & 
\sigma_y=\left[
\begin{array}{cc}
0 &-i \\
i & 0
\end{array}
\right] \ \
&\ , & 
\sigma_z=\left[
\begin{array}{cc}
1 &0 \\
0 & -1
\end{array}
\right]\ \ 
\end{align*}
which act on the base states  $\left|0\right>=
\left[
\begin{array}{c}
0 \\
1 
\end{array}
\right]$ 
and 
$\left|1\right>=
\left[
\begin{array}{c}
1 \\
0 
\end{array}
\right].$
For an initial state consisting of the many qubits (perhaps realized 
by many particles capable of being in two distinct quantum states), such
as $\left|{i_1}{i_2}{i_3}\cdots{i_N}\right>$, a $2N$
dimensional unitary transformation would be implemented to carry
out one step of a computation.

If noise or other spurious interactions occur in the system, quantum coherence may be degraded or lost, and there will be both
a `coherence time' and a `coherence length' over
which the system retains a semblance of its coherence.
A fault-tolerant quantum computer uses 
states that have long coherence times, 
quantum entangled states with long life times, and/or 
error correcting schemes. Systems for transferring qubits
over long distances require long coherence lengths.\footnote{Transferring
qubits across space was first described by C. H. Bennett {\it et al.} in ``Teleporting an Unknown Quantum State via Dual Classical and Einstein-Podolsky-Rosen Channels'', {\it Phys.Rev. Lett.}\,{\bf 70}, pp.\,1895-1899 (1993). Note that transferring a qubit from one system to another does not
violate the no-cloning theorem, because the initial qubit is destroyed in the
process, and that the transfer is cannot be superluminal, as two classical bits
must be sent from the first system to the second before reconstruction of the qubit can take place.}

A new measurement acting on a quantum state 
generally causes some decoherence, so that a number of
components of the wave function may have their phase become stochastically indeterminate.  The
observation of the state of a particle in a multi-particle 
entangled state removes the entanglement of that particle.
As we have seen, measurement of an observable is the equivalent
of projecting out a subspace of the initial state: $\psi_o=P_o\psi$.
Such a projection into a proper subspace is irreversible and non-unitary.
The resulting state of the system no longer
holds information about the complement $(1-P_o)\psi$ state.

In quantum theory, all processes within an isolated system preserve the condition that
the probability of finding any of the possible states of the system
add to unity.  Formally, quantum states evolve by a unitary
transformation.  In the `Copenhagen' view, the act of measurement causes  the wave function for the system to `collapse'.  
A collapse of a quantum state from a 
superposition of substates to one such substate violates unitarity, and therefore is outside the formalism of quantum theory. 
This produces a paradox:  The
measuring instrument is also a physical system, so that the
larger system that contains the observed system and the measuring
devices, left unobserved, should evolve by a unitary transformation,
and no wave function `collapse' should occur.  There is no
easy way out of this paradox.

The measuring devices in the larger combined system must
introduce interactions that do not project out quantum substates in
the combined system, 
but rather redistribute the amplitudes
for various quantum states, making the observed state
highly probable, and the other possible states in the observed system
left with very small amplitudes.  Being unitary for the
combined system of the observed and the measuring 
device, such a measurement
process is, in principle, `reversible'.  The entanglement of
an observed system with a measurement instrument
and subsequent restoration of the original
quantum state has been demonstrated for
simple systems and measuring devices with highly 
restricted interactions.\footnote{See, for example, Nadav Katz, {\it et al.}, ``Reversal of the Weak Measurement of a Quantum State in a Superconducting Phase Qubit'', {\it Phys.Rev.Lett.}\,{\bf 101},\,200401 (2008).} But for
a multitude of interactions, 
restoration after interactions is typically unfeasible with our current resources.\footnote{
This difficulty is related to the ergotic hypothesis in classical mechanics, and the
development of entropy concept in statistical thermodynamics.}
It is also possible that nature does
not just scatter information so much that we cannot easily put systems such as  broken eggs
back together again, but rather actually does lose information over time. 
This possibility is outside the realm of quantum theory.

The same ideas apply to quantum computers. In quantum theory, 
even the measurement of a final state 
after a computer calculation is a reversible process for the computer, 
the measurement device and the surrounding interacting systems.  In principle, no information is lost.  But if the information transferred by erasing a quantum memory state 
produces heat in the environment, some information is practically lost.

If one of the particles in an entangled state is sent to
a second observer as a form of communication, then attempts to intercept 
that particle will degrade or destroy the entanglement, and therefore will be detectable.  This opens the possibility of `absolute' security in transmission lines, particularly since macroscopically long coherence lengths have
been realized with laser beams.

If a set of identical particles are restricted to a two-dimensional
surface, or the space is not simply connected, the quantum state representing two particles may gain a 
phase factor of $\exp(2 i\pi p)$ when the two particles are exchanged, where $p$ need not 
be integer or half-integer.\footnote{By contrast, in a connected region of three dimensional space, there is a space transformation
that will `untangle' the pair, and make $p$ an integer multiple of $1/2$.} If the phase factor $p$ is not $n/2$ (where $n$ is integer),
the particles are called `anyons'.  For three particles, if the order in which the particles
are exchanged produces a different wave function phase, the
group of such exchanges is non-abelian.  This consideration
may be important in the construction of quantum computers through
the storing of information in the topological braiding of non-abelian anyons as they 
progress in space-time.

Topological structures have been shown to be important
in quantum theory.  For example, the continuity condition for the wave function
describing particles adds significance 
to global space-time topology. In some models, particle charges come from
topological structures.  A variety of promising systems for
quantum computers take advantage of the difficulty
of breaking topological structures in order to preserve
quantum coherence, making the system more
`fault immune' against the effect of noise and
other external interactions.

\section{Limits to computing}

\subsection{Practical limits to computing and information storage}

Classical computers have practical limitations in density.  Gates and memory elements smaller
than nanoscale will suffer quantum fluctuations,
with growing uncertainties in bit structures and Boolean transformations as the size of the elements are reduced.  Even our DNA code can be mutated by quantum tunneling.  If the system
has a certain level of noise, classical correction schemes can eliminate errors, at a cost of size.
The techniques to control heat buildup also require volume in the ancillary heat sinks or channels
for radiative cooling. Taking systems at the nanoscale and finding technology that
minimizes heat production toward the Szil{\'a}rd value of $kTln{2}$ per bit lost gives an upper limit to 
computer density made from materials.  Memory and gates based on information in 
light beams have corresponding limits due to pulse duration and wave length uncertainties.

Quantum computers require coherence within the involved quantum states of the computer during computation.
Working against us are physical limitations. For example, the quantum states being used to store information typically have finite lifetimes through spontaneous decay, resulting in the collapse of the employed coherent states.  
Uncontrollable interactions both within and from the outside a quantum computer will tend to collapse coherent states. After sufficient time, coupling to the environment will cause decoherence and disentanglement within a quantum system.

Coherence can be maintained for some period of time by using quantum states which have some intrinsic stability and suffer little debilitating interactions with adjacent systems or with the environment. 
Explorations to find strategies which minimize the limitations are ongoing.
Evidently, each quantum gate must act within the shortest coherence time.  Some
mixing and degradation in quantum states can be tolerated by using repeated calculations and/or implementing error corrections which can reconstruct, with some assurance, a degraded quantum state.  Overall, even though we can anticipate severe practical difficulties to building a quantum computer which can outperform its classical cousin, we see no fundamental limitation, unless our ambitions reach across the cosmos.

\subsection{Cosmological limits}
Strong gravitational fields exist near black holes, which are predicted by
Einstein's General Theory of Relativity to occur when the density of 
an object of mass $m$ exceeds about $3c^6/(2^5\pi G^3m^2)$.  Such black holes got their name because no form of radiation can escape from the hole if it starts out within a region around the hole bounded by a surface called the `horizon'.  For a non-spinning hole without charge, this surface has the `Schwarzschild radius'\,\footnote{Karl Schwarzschild, ``\"Uber das Gravitationsfeld eines Massenpunktes nach der Einsteinschen Theorie'', {\it Sitzungsberichte der K\"oniglich Preussischen Akademie der Wissenschaften}\,{\bf1},\,pp.\,189-196 (1916).} 
$R_S\equiv{2Gm/c^2}$.  Astronomers have found stellar-mass black holes in binary systems by analyzing the orbits of companion stars. Nearby large galaxies are known to contain one or more super-massive black holes at their center, and we suspect all large galaxies do.

Using quantum theory, Hawking showed\footnote{S.W. Hawking, ``Black hole explosions?'', {\it Nature}\,{\bf 248 (5443)},\,pp.30-31 (1976)} that the fluctuations in particle fields near but outside the horizon of a black hole can produce particle pairs with  some of the positive energy particles having sufficient kinetic energy
to reach large distances away, while the negative
energy particles fall into the black hole. Thus, quantum theory requires that black holes evaporate, with a mass loss rate inversely proportional to the square of the hole mass $m$ $\left[dm/dt=-\hbar c^4/(3\cdot 5\cdot 2^{10} \pi G^2\,m^2)\right]$.  The flux of photons emitted is close to that of a hot body at a temperature inversely proportional to $m$ $\left[T=\hbar c^3/(8\pi k Gm)\right]$.

However, to be consistent with quantum theory, a system initially containing  an object and a black hole, with the object destined to disappear into the black hole, with no other interaction but gravity, cannot lose information: The quantum state of the hole and the object evolves unitarily.
One resolution of this paradox is to have the object's information transferred to a region close to the horizon of the black hole.\footnote{It is even possible that the volume surrounded by a black hole horizon is completely empty, even of any space-time structure, with any infalling matter ending up just outside the horizon.}   In this way, Hawking radiation can carry the stored information back out (so the radiation is not perfectly thermal).  Even before Hawking proposed that black holes evaporate, Jacob Berkenstein\footnote{ J. D. Bekenstein, ``Black holes and entropy'', {\it Phys.Rev.}\,{\bf D 7},\,pp.\,2333-2346 (1973).} conjectured that the entropy of a black hole, which is also the information storage capacity, is proportional to the area of the hole's horizon, $4\pi R_S^2$, and inversely proportional to the square of Planck's length. Hawking then calculated the proportionality constant to be $k/4$, where $k$ is Boltzmann's constant.

General Relativity limits the density of a computer, and concurrently the density of information storage.  As a computer becomes larger in a given volume, its density eventually forces the computer to collapse into a black hole.  This leads to the idea that the limiting density of information storage may be effectively two dimensional, with each bit stored in a Planck-size area.  Some (as yet untested) theories even have the information of the whole Universe reflected by a kind of holographic image in one less dimension.

A cosmological limitation on computation also comes from the fact that we appear to live in a finite Universe.  A computer can be no larger than the Universe itself.  Any smaller computer cannot hold the data of the Universe at one time, which is needed to unambiguously project the Universe's future.  In addition, being that the computer is within the Universe, it cannot predict both itself and the Universe.  Our current theories do not incorporate these kinds of limitations, although there are propositions that connect the very small to the very large.

\section{Conclusions}

Quantum computers take advantage of quantum operations in physical systems in order to solve well-posed problems. Quantum theory describes these operations 
based on how nature processes information. Space and time are important primitives in quantum theory, and active participants in both information transfer and information storage.  
While we formulate how nature
handles information, we should recognize that our
physical theories are always tentative. Each covers a limited realm and has a limited accuracy.  
Also, since each theory has a variety of equivalent formulations, with their own language, our main focus should be on the predictions of a theory.
 Even though very successful, quantum theory makes some rather non-intuitive and thought-provoking predictions.  Correspondingly, there are a variety of precautions to which we should be attentive when applying and interpreting the theory.
Reflecting on the underlying ideas central to quantum theory should help us in the exploration of possibilities for future quantum computers.  

\end{document}